%% file: main.tex
\documentclass[10pt,journal,twoside]{IEEEtran}

\usepackage{xcolor}

\usepackage{cite}
\usepackage{amsmath,amssymb,amsfonts}
\usepackage{graphicx}
\usepackage{textcomp}
\usepackage{color}
\usepackage{xcolor}
\usepackage{makecell}
\usepackage{pifont}
\usepackage{wrapfig}
\usepackage{multirow}
\usepackage{booktabs}
\usepackage{caption}
\usepackage{algorithm}
\usepackage{algpseudocode}
\usepackage{comment}
\usepackage{tikz}
\usepackage{pgfplots}
\pgfplotsset{compat=1.18}
\usepackage{enumitem}
\usepackage[table]{xcolor}

\usepackage{hyperref}
\hypersetup{hidelinks=true}

\newcommand{\etal}{\emph{et al. }}

\newcommand{\tick}{\ding{51}}
\newcommand{\cross}{\ding{55}}

\begin{document}
\title{Asynchronous Federated Continual Segmentation with Evolving Clients and Label Spaces}
\author{
\parbox{0.98\textwidth}{\centering\small
Can~Peng\textsuperscript{a}, 
Qianhui~Men\textsuperscript{a,b}, 
Pramit~Saha\textsuperscript{a},  
Qianye~Yang\textsuperscript{a},
Yingyu~Yang\textsuperscript{a},
Shuwei~Xing\textsuperscript{a},
Cheng~Ouyang\textsuperscript{a},
and J.~Alison~Noble\textsuperscript{a}\\
\textsuperscript{a}{University of Oxford, Oxford, United Kingdom}\
\textsuperscript{b}{University of Bristol, Bristol, United Kingdom}
}
\thanks{
\noindent 
Corresponding author: Can Peng (e-mail: can.peng@eng.ox.ac.uk).\\
}
}
\maketitle

\begin{abstract}
    \input{section/0_abstract.tex}
\end{abstract}


\input{section/1_introduction.tex}

\input{section/2_related_work.tex}

\input{section/3_motivation.tex}

\input{section/4_methodology.tex}

\input{section/5_experiment.tex}
\input{section/6_conclusion.tex}

\section*{Acknowledgment}
This work was supported by the UKRI grant EP/X040186/1 (Turing AI Fellowship).
This work was also partly supported by the InnoHK-funded Hong Kong Centre for Cerebro-cardiovascular Health Engineering (COCHE) Project 2.1 (Cardiovascular risks in early life and fetal echocardiography).
PS is supported by the UK EPSRC (Engineering and Physical Research Council) Programme Grant EP/T028572/1 (VisualAI), a UK EPSRC Doctoral Training Partnership award.

\clearpage
\bibliographystyle{reference_style/IEEEtran}
\bibliography{refs}

\clearpage
\appendices
\input{section/7_supplymentary}

\end{document}

%% file: section/0_abstract.tex
Federated learning seeks to foster collaboration among distributed clients while preserving the privacy of their local data.
Traditional federated learning methods typically assume a fixed setting, where participating clients, client data, and learning objectives remain unchanged.
However, in real-world scenarios, a federation may evolve over time, with changes in both its client composition and target label space.
In this evolving federated setting, conventional round-wise model aggregation becomes inflexible, as each federation update requires repeated communication, repeated local computation, and synchronized participation from all accumulated clients.
To address this limitation, we propose CA-MMDS, a continual multiple-model distillation framework for federated continual segmentation with asynchronous clients and evolving label spaces.
Instead of repeatedly aggregating model parameters from all clients, CA-MMDS maintains a server-side archive of client models and updates the global model through proxy-based distillation from multiple archived local models.
When new clients join or existing clients evolve, only the newly added or updated local models need to be uploaded, while unchanged clients can remain offline and continue to contribute through their archived models.
This design substantially reduces communication and computation costs while enabling flexible asynchronous cooperation among evolving clients.
Using multi-class 3D abdominal CT segmentation as an application task, we demonstrate that CA-MMDS efficiently incorporates evolving client knowledge while achieving competitive segmentation performance.

%% file: section/1_introduction.tex
\section{Introduction}
\begin{figure}[t]
    \centering
    \includegraphics[width=1\linewidth, trim={0cm 0cm 0cm 0cm},clip]{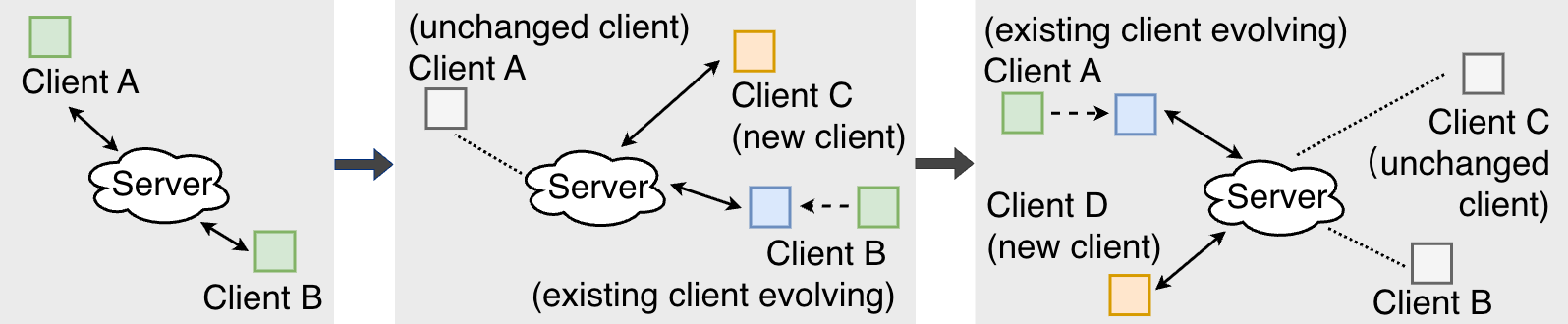} 
    \caption{
    Federated continual segmentation with evolving clients and label spaces, where clients may join or evolve over stages.
    }
    \label{fig1: FCL_pipeline}
    \vspace{-15pt}
\end{figure}

Federated learning (FL) provides a promising framework for collaborative data analysis where data are distributed across institutions and cannot be directly shared due to privacy, legal, or governance constraints~\cite{eden2025scoping}.
Most existing medical FL methods~\cite{xu2023federated,kim2024federated,kim2025communication}, however, are developed under a fixed-federation assumption, where participating clients, local data distributions, and target label spaces are predefined before training and remain unchanged throughout federated optimization.
This assumption is often unrealistic in clinical research and multi-institutional collaborations.
As collaborations evolve, new hospitals or research sites may join the federation, existing sites may collect additional data, and annotation protocols may be extended to include new anatomical or pathological structures.
Figure~\ref{fig1: FCL_pipeline} illustrates such a federated continual learning (FCL) scenario.
In this paper, we study this setting as \textbf{asynchronous federated continual segmentation}, focusing on multi-class 3D medical image segmentation with evolving clients and label spaces.
This setting presents three coupled challenges.
(1) \textbf{Stage-wise federation evolution.}
The federation evolves over stages, requiring the global model to incorporate knowledge from newly introduced or updated clients.
(2) \textbf{Heterogeneous label spaces.}
Clients may have partially overlapping or mismatched label spaces, since datasets are often annotated according to site-specific clinical or research interests.
(3) \textbf{Limited access to historical clients and data.}
Unchanged clients and historical data may become inaccessible due to governance restrictions, data retention policies, or practical coordination costs, making repeated all-client synchronization impractical.
Together, these challenges call for a more flexible FL strategy than conventional fixed-federation training.

The dominant client-server communication paradigm in FL follows model aggregation per communication round (MAPCR) ~\cite{mcmahan2017communication}.
In each round, a global model is distributed to clients, locally updated, and then aggregated at the server.
Although widely adopted, this paradigm is not well suited to evolving FCL scenarios.
Whenever the federation is updated, MAPCR typically requires multi-round training involving all accumulated clients, even when most clients have no new data or revised objectives. 
This leads to repeated communication, computation, and synchronization overhead, and becomes impractical when historical data are inaccessible or unchanged clients cannot be repeatedly brought online.
These limitations motivate a shift from repeated parameter aggregation to server-side knowledge distillation (KD), where client knowledge can be transferred from uploaded local models to a global model using proxy data. 
In this paradigm, each client first trains a local model on its own annotated data and uploads the trained model to the server. 
The server then trains a global model on a proxy dataset by matching its predictions to the aggregated predictions of the uploaded local models.

The most closely related work~\cite{kim2025communication} uses server-side KD with public and batch normalization (BN)-statistics-based synthetic proxy data to reduce communication costs in federated medical segmentation. 
However, it is designed for a fixed-federation setting and assumes that each local model targets a single organ, making the local models binary segmentation models. 
A more realistic medical FL setting should account for local models with multiple, heterogeneous, and potentially overlapping annotation spaces, reflecting site-specific clinical or research interests. 
Since BN statistics are inherently model-level and class-agnostic, BN-statistics-based synthesis becomes less suitable for generating label-conditioned proxy data when local models contain multiple label spaces.
Moreover, directly applying multiple-model distillation with uniform weighting of all source models does not address how archived, newly introduced, and updated client models should be integrated across continual stages, nor does it account for the varying output quality and contributions of different source models. 
The FCL setting therefore requires a stage- and reliability-aware distillation strategy that can distinguish the contributions of different local models during continual aggregation.

To meet this need, we propose \textbf{Continual Archive-based Multiple-Model Distillation in the Server (CA-MMDS)}, an archived-model distillation framework for asynchronous federated continual 3D segmentation. 
CA-MMDS maintains a server-side archive of client models and updates the global model through reliability-weighted distillation on a public proxy dataset, allowing client knowledge to be reused even when unchanged clients remain offline.
For each class and distillation sample, the server adaptively weights local model predictions according to class ownership, model freshness, prediction confidence, and domain compatibility before aggregating them into pseudo-labels. 
These factors reflect practical clinical heterogeneity. 
For example, when generating a pseudo-label for an organ or pathology-related structure in an abdominal CT scan, a recent model trained with annotations for that class and similar contrast-enhanced scans should be trusted more than older models, models without that structure in their label space, or models trained on less related imaging distributions. 
This design enables the server to exploit archived client knowledge while reducing the influence of unreliable, stale, or less relevant predictions during continual federation updates.
The key contributions of this paper are:
\begin{itemize}[leftmargin=10pt]
\item We formulate asynchronous federated continual 3D segmentation as an evolving-client and evolving-label-space problem, where new clients may join and existing clients may expand their annotation targets over time.
\item We propose CA-MMDS, a continual archive-based multiple-model distillation framework that reuses archived client models to update the global model without requiring unchanged clients to return online.
\item We introduce reliability-weighted aggregation for continual distillation, enabling adaptive integration of local models with heterogeneous behaviors and label spaces.
\item We evaluate CA-MMDS on six public abdominal CT segmentation datasets covering 19 anatomical and pathological structures, demonstrating its effectiveness and efficiency.
\end{itemize}

\input{table/FL_method_compare}

%% file: table/FL_method_compare.tex
\definecolor{stepblue}{RGB}{230,242,250} 

\begin{table*}[t]
    \centering
    \caption{
        Comparison of representative FL methods with ours.
        FL challenges addressed: non-iid data (NIID), partially labelled datasets (PL), model heterogeneity (MH), communication efficiency (CE), personalization (P), and continual learning (CL).
        Symbols: $\omega$ model parameters, $z$ logit vectors (model output before softmax), $\Bar{z}$ class-wise average logit vectors, $y$ labels of local data, $\tilde{y}_p$ soft targets (model output after softmax) on public data, $H$ intermediate feature maps,  $A$ attention maps, $D_p$ public/proxy data, $D_s$ synthetic data.
        MAPCR-free indicates whether repeated model aggregation per communication round is not required as the main training mechanism.
        Regularizer denotes using KD as an auxiliary training regularization loss, while digestion denotes using KD as the main supervision for training the global model.
    }
    \footnotesize
    \setlength{\tabcolsep}{5.5pt} 
    \renewcommand{\arraystretch}{1.0} 
    \begin{tabular}{l c c c c c c c}
    \hline
    \multirow{2}{*}{Method} & \multirow{2}{*}{Task} & \multirow{2}{*}{\makecell[c]{Mainly focused\\challenges}} & \multicolumn{2}{c}{Exchanged information} & \multirow{2}{*}{\makecell[c]{MAPCR-\\free?}} & \multicolumn{2}{c}{KD approach} \\
    \cline{4-5} \cline{7-8}
    &  &  & \multicolumn{1}{c}{Sent} & \multicolumn{1}{c}{Received} & & Client-side & Server-side \\
    \hline
    FAug~\cite{jeong2018communication} & Cls & NIID, CE & $\Bar{z}$ & $\Bar{z}$ $+$ generator & \cross & regularizer $+$ generator & generator \\
    FedDF~\cite{lin2020ensemble} & Cls & NIID, MH & $\omega$ & $\omega$ & \cross & - & digestion $+$ $D_p$  \\
    CFD~\cite{sattler2021cfd} & Cls & MH, CE & $\tilde{y}_p$ & $\tilde{y}_p$ & \cross & digestion $+$ $D_p$ & digestion $+$ $D_p$ \\
    FedAD~\cite{gong2021ensemble} & Cls & MH, CE & $z_p, A_p$ & - & \tick & - & digestion $+$ $D_p$  \\
    FedGKT~\cite{he2020group} & Cls & MH, CE & $z, H, y$ & $z$ & \cross & regularizer & regularizer \\
    Chen \etal~\cite{chen2023spectral} & Cls & P & $\omega$ & $\omega$ & \cross & regularizer & - \\
    Target~\cite{zhang2023target}& Cls & CL & $\omega$ & $\omega + D_s$ & \cross & regularizer & generator \\
    Re-Fed~\cite{li2024towards} & Cls & CL & $\omega$ & $\omega$ & \cross & - & - \\
    CFeD~\cite{ma2022continual} & Cls & NIID, CL & $\omega$ & $\omega$ & \cross & digestion $+$ $D_p$ & digestion $+$ $D_p$\\
    Fed-MENU~\cite{xu2023federated} & Seg & NIID, PL & $\omega$ & $\omega$ & \cross & - & - \\
    Kim \etal~\cite{kim2024federated} & Seg & NIID, PL & $\omega$ & $\omega$ & \cross & regularizer & - \\
    Kim \etal~\cite{kim2025communication} & Seg & NIID, PL, CE & $\omega$ & $\omega$ & \tick & - & digestion $+$ $D_p$ $+$ $D_s$ \\
    \rowcolor{stepblue} Ours & Seg & NIID, PL, CE, CL & $\omega$ & $\omega$ & \tick & - & reliability-weighted digestion $+$ $D_p$ \\
    \hline
    \end{tabular}
    \label{tab:KD_in_FL}
    \vspace{-5pt}
\end{table*}

%% file: section/2_related_work.tex
\section{Related Work}

\subsection{Federated Continual Learning (FCL)}
Most FL frameworks assume a fixed federation with predefined clients, data distributions, and learning objectives.
These settings mainly address spatial heterogeneity across clients, such as non-IID data distributions, while often overlooking temporal heterogeneity arising from the evolution of the federation.
Recently, FCL has been explored to enable a global model to adapt to changing client data and task requirements over time~\cite{usmanova2021distillation,dong2022federated,ma2022continual,qi2023better,tran2024text,babakniya2024data,zhang2023target,li2024towards}.
Existing FCL methods commonly tackle this challenge through synthetic data replay~\cite{dong2022federated,qi2023better,tran2024text,babakniya2024data,zhang2023target} or memory replay with locally retained exemplars~\cite{usmanova2021distillation,li2024towards}.
However, current FCL studies mainly focus on image classification, leaving federated continual segmentation underexplored.
Compared with FCL classification, where each sample is assigned a single class label, FCL segmentation is more challenging because it requires dense, spatially structured predictions.
In this work, we focus on FCL for 3D medical image segmentation with evolving clients and label spaces.

\subsection{Federated Medical Image Segmentation}
Federated medical image segmentation has been studied under various forms of heterogeneity, including distribution shift~\cite{liu2021feddg,xu2022closing,wang2022personalizing,wang2023feddp}, weak annotations~\cite{zhu2023feddm,wicaksana2022fedmix}, and noisy labels~\cite{wu2024feda3i}.
While many studies focus on binary segmentation, federated multi-class segmentation remains less explored despite its practical importance.
This paper targets 3D abdominal CT segmentation, where different institutions may partially annotate different anatomical or pathological structures due to differences in clinical expertise, annotation protocols, and local task priorities.
Several recent works have investigated the fixed-FL formulation of this task.
Xu \etal proposed class-specific encoders to improve feature extraction under inconsistent labels~\cite{xu2023federated}.
Kim \etal used KD to regularize local training and mitigate prediction forgetting caused by partial annotations~\cite{kim2024federated}.
More recently, Kim \etal proposed a communication-efficient FL framework that distills local models at the server using public and synthetic images~\cite{kim2025communication}.
These methods address challenges such as inconsistent labels, partial annotations, and communication efficiency.
In contrast, our work focuses on asynchronous federated continual segmentation, where new clients may join and existing clients may revise their label spaces over multiple stages.
This setting requires the server to continually update the global model as the federation evolves, while preserving previously acquired knowledge without accessing historical data or requiring unchanged clients to participate.

\subsection{Knowledge Distillation (KD) in FL}
KD~\cite{hinton2015distilling} has been widely adopted in FL, as it enables knowledge transfer through input-output mappings without requiring access to the original training data.
Existing KD-based FL methods use distillation for various purposes, including improving communication efficiency~\cite{jeong2018communication,sattler2021cfd,kim2025communication}, alleviating model heterogeneity~\cite{lin2020ensemble,afonin2021towards,gong2021ensemble}, supporting personalization~\cite{chen2023spectral}, and mitigating forgetting in FCL~\cite{ma2022continual,zhang2023target,li2024towards}.
Table~\ref{tab:KD_in_FL} summarizes representative KD-based FL methods and compares them with our approach\footnote{For a survey covering a broader scope, refer to~\cite{mora2024knowledge}.}.
Although KD provides a flexible mechanism for transferring knowledge across distributed models, most existing KD-based FL methods are not designed for FCL segmentation.
Many methods still rely on model aggregation per communication round (MAPCR), which requires repeated synchronization with clients.
Communication-efficient KD methods such as~\cite{kim2025communication} reduce this dependency, but they are typically developed for fixed-federation settings.
In such settings, distillation is performed over a fixed set of local models, which are usually treated equally when transferring knowledge to the global model.
This does not directly address FCL, where archived, newly introduced, and updated client models may coexist across stages, and where local models may have heterogeneous and evolving label spaces.
By contrast, CA-MMDS targets this continual setting by maintaining an archive of local models and introducing reliability-weighted aggregation to adaptively integrate local model knowledge during continual federation updates.

%% file: section/3_motivation.tex
\begin{figure*}[t]
    \vspace{-10pt}
    \centering
    \includegraphics[width=0.95\linewidth, trim={0cm 0cm 0cm 0cm},clip]{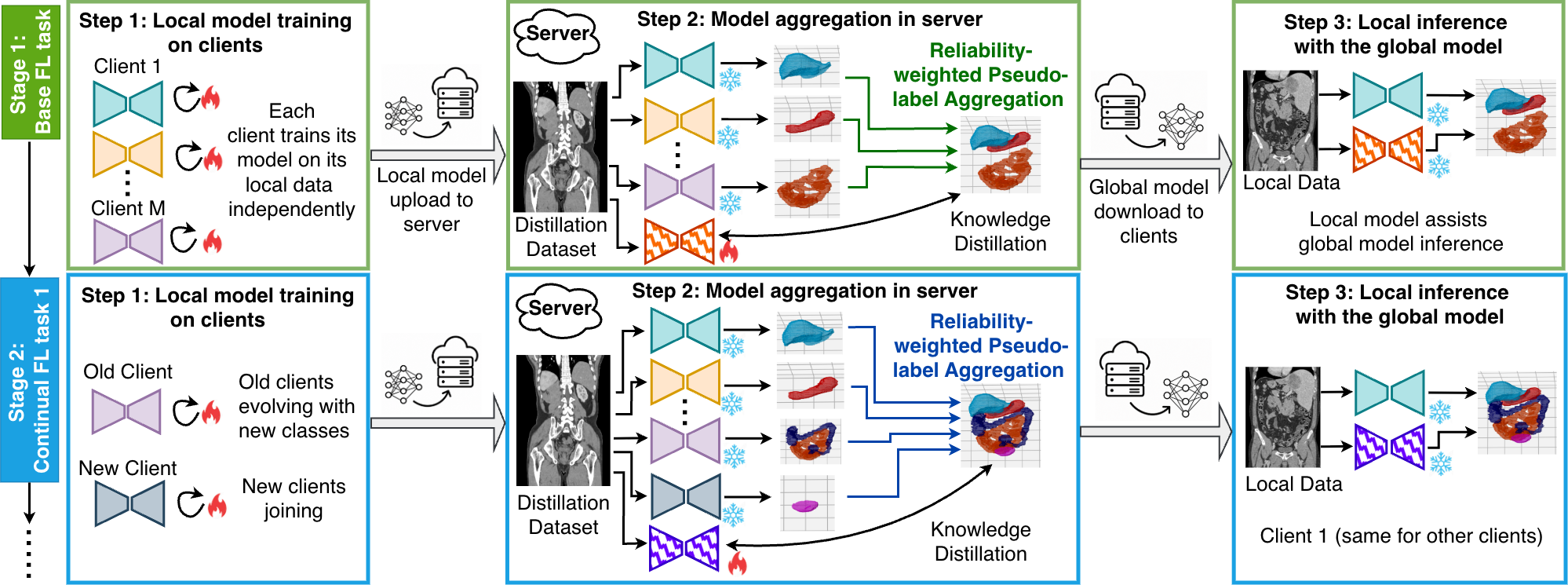} 
    \caption{
    Overview of the proposed CA-MMDS framework.
    The server archives uploaded local models, performs reliability-weighted multi-model distillation on public proxy data, and updates the global model as new or revised clients become available, while unchanged clients remain offline.
    }
    \vspace{-10pt}
	\label{fig1: framework}
\end{figure*}

\section{Problem Formulation}
\label{Problem Formulation}

We consider FCL over $T$ stages.
At each stage $t \in \{1,\ldots,T\}$, the federation is characterized by a cumulative client set $\mathcal{S}_t$ and a cumulative label set $\mathcal{C}_t$.
Each client $k \in \mathcal{S}_t$ owns a private dataset $\mathcal{D}_{k,t}$ with annotations for a client-specific label set $\mathcal{C}_{k,t} \subseteq \mathcal{C}_t$.
Due to site-specific annotation protocols and evolving clinical or research interests, local label sets may differ across clients, partially overlap, or expand over time.
For example, one client may focus on kidney-related structures, such as \{\texttt{kidney, kidney stone, kidney tumor}\}, while another may focus on liver-related structures, such as \{\texttt{liver, liver tumor}\}.
Meanwhile, multiple clients may share common targets, such as frequently annotated abdominal organs.
The federation evolves through two types of updates.
First, new clients may join the federation with their local datasets and annotation targets.
Second, existing clients may update their local models when additional data become available or when their annotation targets are extended.
Let $\mathcal{U}_t \subseteq \mathcal{S}_t$ denote the set of clients that are newly added or updated at stage $t$.
Clients in $\mathcal{S}_t \setminus \mathcal{U}_t$ have no new data or objectives at this stage and are assumed to be unavailable for retraining or synchronized communication.
The goal at each stage $t$ is to obtain a global segmentation model that can segment the cumulative label set
$
\mathcal{C}_t = \bigcup_{k \in \mathcal{S}_t} \mathcal{C}_{k,t},
$
without sharing raw client data with other clients or with the server.

%% file: section/4_methodology.tex
\section{Method}

We propose \textbf{Continual Archive-based Multiple-Model Distillation in the Server (CA-MMDS)}, a KD-based asynchronous FCL segmentation framework.
At each stage, only newly added or updated clients train local models and upload them to the server, while unchanged clients remain offline and are represented by their archived models.
Given a public unlabeled proxy dataset, the server collects predictions from the archived local models.
Since these models may differ in class ownership, prediction confidence, freshness, and domain compatibility, CA-MMDS performs reliability-weighted aggregation to generate class-wise pseudo-labels while reducing the influence of unreliable, stale, or less relevant predictions.
The pseudo-labels are then used to train an updated global segmentation model.
The framework is illustrated in Figure~\ref{fig1: framework}.

\subsection{Local Model Training}
At each stage $t$, only clients in $\mathcal{U}_t$ perform local training.
Each client $k \in \mathcal{U}_t$ trains a segmentation model $f^L_{k,t}$ on its private dataset $\mathcal{D}_{k,t}$, where the superscript $L$ distinguishes local models from the global model $f^G_t$ maintained by the server.
To support heterogeneous and expanding label sets across clients, we follow~\cite{liu2023clip} and use CLIP text embeddings~\cite{radford2021learning} to represent class semantics, allowing new classes to be introduced through class prompts without modifying the network architecture.
For a 3D input volume $\mathbf{X} \in \mathbb{R}^{C_{\mathrm{in}}\times H \times W \times Z}$ sampled from $\mathcal{D}_{k,t}$, the local model predicts:
{
\begin{equation}
\mathbf{P}^L_{k,t} = f^L_{k,t}(\mathbf{X}) \in [0,1]^{|\mathcal{C}_{k,t}|\times H\times W\times Z},
\label{eq:local_prediction}
\end{equation}
}
where $C_{\mathrm{in}}$ denotes the number of input channels, and each output channel $\mathbf{P}^L_{k,t,c}$ corresponds to a class $c \in \mathcal{C}_{k,t}$.
We denote the spatial voxel domain of $\mathbf{X}$ by $\Omega_{\mathbf{X}}$, and use $n\in\Omega_{\mathbf{X}}$ to index voxel locations.

Let $\mathbf{Y}_{k,t,c}$ denote the binary ground-truth mask for class $c$.
Since each client only annotates its own label set, the local objective is computed over $\mathcal{C}_{k,t}$:
{
\begin{equation}
\begin{aligned}
\mathcal{L}^{L}_{k,t} =\frac{1}{|\mathcal{C}_{k,t}|}\sum_{c \in \mathcal{C}_{k,t}}
\Big[
&\mathcal{L}_{\mathrm{BCE}}\left(\mathbf{P}^L_{k,t,c},\mathbf{Y}_{k,t,c}\right) \\
&+ \mathcal{L}_{\mathrm{Dice}}\left(\mathbf{P}^L_{k,t,c},\mathbf{Y}_{k,t,c}\right)
\Big].
\end{aligned}
\label{eq:local_training_loss}
\end{equation}
}
After training, client $k$ uploads $f^L_{k,t}$ and $\mathcal{C}_{k,t}$ to the server.

\subsection{Continual Model Archiving}
The server maintains a model archive that stores all local models uploaded by clients throughout the federation process. 
Instead of requiring all clients to participate at every federation stage, unchanged clients can remain offline because their archived models represent their local knowledge during server-side distillation.
At stage $t$, the archive is defined as
{
\begin{equation}
\mathcal{A}_t =\left\{\left(f^L_{k,\tau}, \mathcal{C}_{k,\tau}, \tau\right) \mid k \in \mathcal{S}_t,\ \tau \in \mathcal{H}_{k,t}
\right\},
\label{eq:model_archive}
\end{equation}
}
where $\mathcal{H}_{k,t} = \{\tau \mid 1 \leq \tau \leq t,\ k \in \mathcal{U}_{\tau}\}$ denotes the set of historical upload stages for client $k$ up to stage $t$.
Each archived entry consists of a local model $f^L_{k,\tau}$, its corresponding label set $\mathcal{C}_{k,\tau}$, and the upload stage $\tau$.
This archive enables CA-MMDS to perform asynchronous federation updates by aggregating knowledge from newly uploaded, revised, and historical local models.

\subsection{Reliability-Weighted Aggregation}
Given the model archive $\mathcal{A}_t$, the server aggregates knowledge from archived local models through reliability-weighted multi-model distillation.
For a proxy volume $\mathbf{X}$ from the public unlabeled distillation dataset $\mathcal{D}_p$, each archived local model $f^L_{k,\tau}$ produces a probability map:
{
\begin{equation}
\mathbf{P}^{L}_{k,\tau} = f^L_{k,\tau}(\mathbf{X}).
\label{eq:archive_prediction}
\end{equation}
}
Directly averaging predictions from archived models is suboptimal, as model reliability may vary across classes and proxy samples due to class ownership, model freshness, prediction confidence, and domain compatibility.
CA-MMDS therefore assigns reliability scores to these model predictions and uses the resulting weights for pseudo-label aggregation.
The following subsections detail these four factors.

\noindent\textbf{Class ownership.}
For each class $c \in \mathcal{C}_t$, we define a class ownership indicator for each archived local model $f^L_{k,\tau}$:
{
\begin{equation}
o_{k,\tau,c} = \mathbb{I} \left[ c \in \mathcal{C}_{k,\tau} \right],
\label{eq:class_ownership}
\end{equation}
}
where $o_{k,\tau,c}=1$ indicates that $f^L_{k,\tau}$ was trained for class $c$.

\noindent\textbf{Model freshness.}
To account for the temporal evolution of the federation, we define a freshness score:
{
\begin{equation}
r_{k,\tau,t} = \exp \left( -\gamma (t - \tau) \right),
\label{eq:freshness_score}
\end{equation}
}
where $\gamma \geq 0$ controls the degree of down-weighting.

\noindent\textbf{Prediction confidence.}
For local models whose label space includes class $c$, we estimate their prediction confidence on the proxy sample using voxel-wise binary entropy over the spatial voxel domain $\Omega_{\mathbf{X}}$:
{
\begin{equation}
\begin{aligned}
E_{k,\tau,c}(\mathbf{X}) 
&= -\frac{1}{|\Omega_{\mathbf{X}}|}\sum_{n\in\Omega_{\mathbf{X}}}\Big[
\mathbf{P}^{L}_{k,\tau,c,n}\log\left(\mathbf{P}^{L}_{k,\tau,c,n}+\epsilon\right) \\
&+\left(1-\mathbf{P}^{L}_{k,\tau,c,n}\right)\log\left(1-\mathbf{P}^{L}_{k,\tau,c,n}+\epsilon\right)\Big].
\end{aligned}
\label{eq:binary_entropy}
\end{equation}
}
where $\mathbf{P}^{L}_{k,\tau,c,n}$ denotes the predicted probability of class $c$ at voxel location $n$.
A lower entropy value indicates a more confident prediction.
We then convert the entropy into a confidence score:
{
\begin{equation}
q_{k,\tau,c}(\mathbf{X}) = \exp \left( -E_{k,\tau,c}(\mathbf{X}) / \beta \right),
\label{eq:confidence_score}
\end{equation}
}
where $\beta > 0$ is a temperature parameter.

\noindent\textbf{Domain compatibility.}
To account for potential domain mismatch between the distillation dataset and the local client data, we introduce a domain compatibility term.
The running statistics stored in BN layers provide a compact summary of the local feature distribution without incurring additional communication cost.
Let
{
\begin{equation}
\mathcal{B}_{k,\tau} = \left\{\left(\boldsymbol{\mu}^{j}_{k,\tau}, \boldsymbol{\sigma}^{j}_{k,\tau}\right) \mid
j \in \mathcal{J}_{\mathrm{BN}}\right\}
\label{eq:bn_statistics}
\end{equation}
}
denote the BN statistics stored in the archived local model $f^L_{k,\tau}$, where $\boldsymbol{\mu}^{j}_{k,\tau}$ and $\boldsymbol{\sigma}^{j}_{k,\tau}$ are the running mean and standard deviation of the $j$-th BN layer, respectively, and $\mathcal{J}_{\mathrm{BN}}$ denotes the set of BN layers.
For a proxy volume $\mathbf{X}$, we forward it through $f^L_{k,\tau}$ and compute the corresponding activation statistics $\widehat{\boldsymbol{\mu}}^{j}_{k,\tau}(\mathbf{X})$ and $\widehat{\boldsymbol{\sigma}}^{j}_{k,\tau}(\mathbf{X})$ for each BN layer $j$.
The BN compatibility distance is defined as
{
\begin{equation}
\begin{aligned}
d_{k,\tau}(\mathbf{X}) = \frac{1}{|\mathcal{J}_{\mathrm{BN}}|}\sum_{j \in \mathcal{J}_{\mathrm{BN}}}\Big(
&\left\|\widehat{\boldsymbol{\mu}}^{j}_{k,\tau}(\mathbf{X})-\boldsymbol{\mu}^{j}_{k,\tau}\right\|_2 \\
&+\left\|\widehat{\boldsymbol{\sigma}}^{j}_{k,\tau}(\mathbf{X})-\boldsymbol{\sigma}^{j}_{k,\tau}\right\|_2\Big).
\end{aligned}
\label{eq:bn_distance}
\end{equation}
}

The corresponding domain compatibility score is
{
\begin{equation}
b_{k,\tau}(\mathbf{X}) = \exp\left(-d_{k,\tau}(\mathbf{X}) / \eta \right),
\label{eq:bn_compatibility}
\end{equation}
}
where $\eta>0$ is a temperature parameter.

For readability, we use $m$ to index an archived local model in $\mathcal{A}_t$ and omit its explicit client and upload-stage indices.
Accordingly, $o_{m,c}$, $r_{m,t}$, $q_{m,c}(\mathbf{X})$, and $b_m(\mathbf{X})$ denote the class ownership, freshness, confidence, and domain compatibility scores of archived model $m$, respectively.
Combining the four factors, the reliability weight of archived model $m$ for class $c$ on proxy volume $\mathbf{X}$ is computed as
{
\begin{equation}
\alpha_{m,c}(\mathbf{X}) = \frac{o_{m,c} r_{m,t} q_{m,c}(\mathbf{X}) b_m(\mathbf{X})}
{
\sum\limits_{m' \in \mathcal{A}_t} o_{m',c} r_{m',t} q_{m',c}(\mathbf{X}) b_{m'}(\mathbf{X}) +\epsilon
},
\label{eq:model_reliability}
\end{equation}
}
where $\epsilon$ is a small constant for numerical stability.
The aggregated pseudo-label for class $c$ is then obtained by reliability-weighted ensembling:
{
\begin{equation}
\widetilde{\mathbf{Y}}_{t,c}(\mathbf{X})=\sum_{m \in \mathcal{A}_t}
\alpha_{m,c}(\mathbf{X})\mathbf{P}^{L}_{m,c}(\mathbf{X}).
\label{eq:aggregated_pseudo_label}
\end{equation}
}

\input{table/pseudo_code}
\input{table/communication_load}

\subsection{Global Model Distillation}
After obtaining the reliability-weighted pseudo-labels, the server trains an updated global segmentation model $f^G_t$ over the cumulative label set $\mathcal{C}_t$.
For each proxy volume $\mathbf{X} \in \mathcal{D}_p$, the global model predicts
{
\begin{equation}
\mathbf{P}^{G}_{t} = f^G_t(\mathbf{X}),
\label{eq:global_prediction}
\end{equation}
}
where $\mathbf{P}^{G}_{t,c}$ denotes the predicted probability map for class $c \in \mathcal{C}_t$.
The aggregated pseudo-label $\widetilde{\mathbf{Y}}_{t,c}(\mathbf{X})$ provides supervision for training the global model.
The distillation objective combines soft binary cross-entropy loss and soft Dice loss:
{
\begin{equation}
\mathcal{L}^{G}_{\mathrm{dist}} = \frac{1}{|\mathcal{C}_t|}
\sum_{c \in \mathcal{C}_t}
\left[\mathcal{L}_{\mathrm{BCE}}\left(\mathbf{P}^{G}_{t,c}, \widetilde{\mathbf{Y}}_{t,c}\right)
+\mathcal{L}_{\mathrm{Dice}}\left(\mathbf{P}^{G}_{t,c}, \widetilde{\mathbf{Y}}_{t,c}\right)\right].
\label{eq:global_distillation}
\end{equation}
}

After distillation, the updated global model $f^G_t$ is redistributed to clients.
The complete pipeline of CA-MMDS is summarized in Algorithm~\ref{pseudo_code: algorithm}.

\subsection{Efficiency Analysis}
CA-MMDS improves efficiency by reusing archived local models, so unchanged clients can remain offline and only newly added or updated clients need to train and upload.
This avoids repeated all-client synchronization during global model updates.
We compare its communication and computation costs with MAPCR in Table~\ref{tab:communication_computation}.

\noindent\textbf{Communication.}
In MAPCR, each federation update requires $E$ rounds of server-client communication to train the global model to convergence.
At stage $t$, both upload and download costs scale as $E|\mathcal{S}_t|$, yielding a total communication load of $2E|\mathcal{S}_t|$.
In contrast, CA-MMDS requires only newly added or updated clients $\mathcal{U}_t$ to upload their local models,  while the server redistributes the updated global model to all accumulated clients $\mathcal{S}_t$, resulting in a total communication load of $|\mathcal{U}_t|+|\mathcal{S}_t|$.
Since $\mathcal{U}_t \subseteq \mathcal{S}_t$ and $E \gg 1$, CA-MMDS substantially reduces communication overhead by avoiding repeated round-based synchronization with all clients.

\noindent\textbf{Computation.}
In MAPCR, all accumulated clients $\mathcal{S}_t$ participate in each federation update, so client-side computation scales as $|\mathcal{S}_t|O$.
In contrast, CA-MMDS performs local training only for newly added or updated clients $\mathcal{U}_t$, while unchanged clients remain offline and are represented by their archived models, reducing client-side computation to $|\mathcal{U}_t|O$.
On the server side, CA-MMDS performs one distillation process, including inference over archived models, reliability-weighted aggregation, and global model training.
As archived-model inference and aggregation are substantially cheaper than full model training, the server-side computation is approximated as $O$, yielding a total computation cost of $(|\mathcal{U}_t|+1)O$.

%% file: table/pseudo_code.tex
\definecolor{stepblue}{RGB}{230,242,250} 

\begin{algorithm}[t]
\footnotesize
\caption{Continual-Aware Multi-Model Distillation in the Server (CA-MMDS)}
\label{pseudo_code: algorithm}

\textbf{Input:} $T$: number of federation stages; 
$\mathcal{U}_t$: newly added or updated clients at stage $t$; 
$\mathcal{S}_t$: accumulated client set at stage $t$; 
$\mathcal{D}_{k,t}$: private local dataset of client $k$ at stage $t$; 
$\mathcal{D}_{p}$: public unlabeled distillation dataset; 
$\mathcal{A}_0=\emptyset$: initial server-side model archive.

\begin{algorithmic}[1]
\For{$t = 1$ \textbf{to} $T$}

    \State \colorbox{stepblue}{\strut\textbf{Step 1: Local training}}
    \For{each client $k \in \mathcal{U}_t$ \textbf{in parallel}}
        \State Train local model $f^L_{k,t}$ on $\mathcal{D}_{k,t}$ using Eq.~\ref{eq:local_prediction}--\ref{eq:local_training_loss}.
        \State Upload $f^L_{k,t}$ and $\mathcal{C}_{k,t}$ to the server.
    \EndFor

    \State \colorbox{stepblue}{\strut\textbf{Step 2: Model archive update}}
    \State Update model archive: $\mathcal{A}_t \leftarrow \mathcal{A}_{t-1} \cup \left\{\left(f^L_{k,t}, \mathcal{C}_{k,t}, t\right) \mid k \in \mathcal{U}_t \right\}$.
    \State Obtain the cumulative label set $\mathcal{C}_t$ from the archived label sets.

    \State \colorbox{stepblue}{\strut\textbf{Step 3: Reliability-weighted pseudo-label aggregation}}
    \For{each archived model $m = (f^L_{m},\mathcal{C}_{m},\tau) \in \mathcal{A}_t$}
        \State Generate local model predictions $\mathbf{P}^{L}_{m}$ on $\mathcal{D}_{p}$ using Eq.~\ref{eq:archive_prediction}.
    \EndFor
    \For{each proxy volume $\mathbf{X} \in \mathcal{D}_{p}$ and each class $c \in \mathcal{C}_t$}
        \For{each archived model $m \in \mathcal{A}_t$}
            \State Compute $o_{m,c}$, $r_{m,t}$, $q_{m,c}(\mathbf{X})$, and $b_{m}(\mathbf{X})$ using Eqs.~\ref{eq:class_ownership}--\ref{eq:bn_compatibility}.
            \State Compute reliability weights $\alpha_{m,c}(\mathbf{X})$ using Eq.~\ref{eq:model_reliability}.
        \EndFor
        \State Aggregate predictions into pseudo-labels $\widetilde{\mathbf{Y}}_{t,c}(\mathbf{X})$ using Eq.~\ref{eq:aggregated_pseudo_label}.
    \EndFor

    \State \colorbox{stepblue}{\strut\textbf{Step 4: Global model distillation}}
    \State Train global model $f^G_t$ on $\mathcal{D}_{p}$ using $\widetilde{\mathbf{Y}}_{t,c}(\mathbf{X})$ and Eq.~\ref{eq:global_prediction}--\ref{eq:global_distillation}.
    \State Distribute $f^G_t$ to clients in $\mathcal{S}_t$.
\EndFor
\end{algorithmic}
\end{algorithm}
\vspace{-5pt}

%% file: table/communication_load.tex
\begin{table*}[t]
    \centering
    \caption{
        Communication and computation comparison between MAPCR and the proposed CA-MMDS.
        $\mathcal{S}_t$ denotes the accumulated client set, and $\mathcal{U}_t$ denotes the newly added or updated clients at stage $t$.
        Each model transmission between the server and one client is counted as one communication.
        In CA-MMDS, only clients in $\mathcal{U}_t$ perform local training, while unchanged clients are represented by archived models on the server.
        Its server-side computation is approximated as $O$ because it mainly consists of one global model distillation process, with lightweight teacher inference and reliability-weighted aggregation.
    }
    \footnotesize
    \renewcommand{\arraystretch}{1}
    \begin{tabular}{l|c c c|c c c|c}
        \hline
        \multicolumn{8}{c}{
        Stage $t$: $\mathcal{U}_t \subseteq \mathcal{S}_t$, $E \gg 1$, and $O$ denotes the cost of training one model to convergence
        } \\
        \hline
        \multirow{2}{*}{Method}
        & \multicolumn{3}{c|}{Communication}
        & \multicolumn{3}{c|}{Computation}
        & \multirow{2}{*}{\makecell{Supports offline\\accumulated clients?}} \\
        \cline{2-7}
        & Upload
        & Download
        & Total
        & Client-side
        & Server-side
        & Total
        & \\
        \hline
        MAPCR 
        & $E|\mathcal{S}_t|$
        & $E|\mathcal{S}_t|$
        & $2E|\mathcal{S}_t|$
        & $|\mathcal{S}_t|O$
        & $\approx 0$
        & $|\mathcal{S}_t|O$
        & \cross \\
        \hline
        CA-MMDS 
        & $|\mathcal{U}_t|$
        & $|\mathcal{S}_t|$
        & $|\mathcal{U}_t|+|\mathcal{S}_t|$
        & $|\mathcal{U}_t|O$
        & $\approx O$
        & $(|\mathcal{U}_t|+1)O$
        & \tick \\
        \hline
    \end{tabular}
    \label{tab:communication_computation}
    \vspace{-5pt}
\end{table*}

%% file: section/5_experiment.tex
\section{Experiment}
\label{Experiment}

\input{section/5_1_experiment_setup}

\input{exp/main_compare_tab}
\input{exp/main_compare}

\subsection{Main Comparison}
\label{subsec: Main Comparison}
Table~\ref{tab:main_comparison} compares CA-MMDS with baselines after five stages of FCL segmentation. 
Centralized training assumes access to all client data and serves as the upper bound, while FedAvg~\cite{mcmahan2017communication} represents conventional multi-round FL. 
We also report FedAvg and FedProx~\cite{li2020federated} with only 1 or 10 communication rounds to examine low-communication MAPCR-style alternatives. 
Although multi-round FedAvg achieves competitive segmentation performance, it requires 38,000 model transmissions and $19\times$ computation, where $1\times$ denotes the cost of training one client model to convergence. 
This makes it difficult to deploy in real-world cross-institutional medical collaborations where clients may differ in data scale, computational resources, and availability. 
However, reducing the number of communication rounds leads to near-failed segmentation performance for MAPCR-style methods, indicating that simply limiting synchronization is insufficient. 
Distillation-based methods avoid repeated round-based aggregation and reduce the communication cost to only 25 transmissions while supporting offline clients through archived local models.
For the adapted Kim et al.-style KD baseline~\cite{kim2025communication}, the additional BN-statistics-based synthesis cost is estimated following its setting, where generating 200 synthetic patches for each uploaded local model corresponds to approximately half of one model-training computation. 
During training, we sample equal numbers of public proxy samples and synthetic proxy samples in each epoch to balance their contributions. 
CA-MMDS improves upon this adapted KD baseline by 1.64\% in overall Dice and reduces overall ASSD by 
7.92 mm, suggesting that class-agnostic BN-statistics-based synthesis provides limited benefit when distilling knowledge from multi-label local segmentation models.
Compared with Uniform MMDS, which uniformly aggregates predictions from local models covering the target class, CA-MMDS further improves overall Dice by 1.22\% and reduces overall ASSD by 2.33 mm. These results demonstrate the benefit of reliability-weighted aggregation in continual server-side distillation.

Figure~\ref{fig3: communication} provides a stage-wise comparison of the communication and computation costs of FedAvg and CA-MMDS. 
As the federation evolves, FedAvg repeatedly communicates with all accumulated clients over multiple rounds, resulting in 38,000 model transmissions by stage 5. 
In contrast, CA-MMDS requires only 25 transmissions, as only newly added or updated clients upload their local models, while unchanged clients are represented by their archived models. 
A similar trend is observed in computation cost: FedAvg accumulates a cost of $19\times$ by stage 5, whereas CA-MMDS requires only $11\times$ by reusing archived client models and avoiding repeated local retraining for unchanged clients. 
The efficiency gap widens over stages, highlighting the scalability and flexibility of CA-MMDS in asynchronous federated continual segmentation.

\input{exp/ablation_exp}
\input{exp/proxy_dataset_analysis_tab}

Figure~\ref{fig4: accuracy} presents the class-wise Dice and ASSD results across the five-stage FCL process. 
For the 15 organ classes, CA-MMDS achieves Dice scores above 65\% for all classes, with 12 classes exceeding 70\%, indicating its ability to effectively aggregate anatomical knowledge from evolving clients and label spaces. 
In contrast, the pathology-related classes remain substantially more challenging. 
This is likely because the proxy dataset WORD~\cite{luo2021word} mainly contains healthy anatomical structures and provides limited pathological patterns, making it difficult to activate and distill pathology-specific knowledge from local model predictions. 
These results show that CA-MMDS achieves competitive performance on organ classes, while also highlighting pathology-aware proxy data or label-conditioned data-free distillation as important directions for further improvement.

\subsection{Ablation Study}
Table~\ref{tab:ablation_reliability} evaluates the contribution of each reliability component in CA-MMDS. Uniform MMDS serves as the baseline without reliability-aware weighting, where we retain the class-ownership constraint for fair comparison but aggregate the predictions of all local models covering the target class with equal weights. 
Compared with Uniform MMDS, CA-MMDS improves Overall Dice by 1.22\% and reduces Overall ASSD by 2.33 mm. 
For organ classes, CA-MMDS improves Dice by 1.28\% and reduces ASSD by 2.58 mm, achieving the best organ performance among all variants. 
For pathology classes, CA-MMDS also improves Dice by 0.95\% and reduces ASSD by 1.37 mm over Uniform MMDS, although pathology segmentation remains more challenging. 
Removing freshness, prediction confidence, or domain compatibility leads to degradation in either Dice or ASSD, indicating that these components provide complementary reliability cues for pseudo-label aggregation. 
These results demonstrate the importance of reliability-aware aggregation in asynchronous FCL with evolving clients and label spaces.

\subsection{Distillation Dataset Analysis}
We further analyse CA-MMDS with respect to the choice and scale of the public proxy dataset used for server-side distillation. 
In addition to the default WORD~\cite{luo2021word} proxy dataset, we consider reduced WORD subsets and a pathology-oriented abdominal CT proxy dataset, HCC-TACE-SEG~\cite{moawad2023multimodality}. WORD provides broad abdominal organ coverage, whereas HCC-TACE-SEG contains 105 patient-level CT volumes from hepatocellular carcinoma cases treated with transarterial chemoembolization.
As shown in Table~\ref{tab:proxy_sensitivity}, the full WORD proxy achieves strong organ segmentation performance, indicating that broad anatomical coverage is beneficial for multi-organ distillation. However, its performance on pathology classes is limited, likely because WORD contains few pathological patterns, as discussed in Section~\ref{subsec: Main Comparison}.
Reducing the WORD proxy scale consistently degrades performance, showing that proxy data scale remains important when transferring knowledge across heterogeneous label spaces.
Compared with WORD, HCC-TACE-SEG yields lower organ performance, likely due to its more limited anatomical diversity. 
However, its pathology-oriented data improve pathology segmentation over WORD by 2.99\% in Dice and reduce ASSD by 2.99 mm. 
This gain is mainly driven by the liver tumor class, whose Dice score increases from 42.47\% to 50.75\%, suggesting that relevant liver pathology in the proxy dataset helps activate and distill liver tumor knowledge from local models. 
Using only 50\% of HCC-TACE-SEG substantially degrades performance, further confirming that both proxy semantics and data scale affect distillation quality.
Combining WORD and HCC-TACE-SEG achieves the best performance for both organ and pathology structures, suggesting that an effective proxy dataset should cover the full semantic label space, including both anatomical organs and pathological structures. 
Interestingly, combining WORD with the full HCC-TACE-SEG dataset does not outperform combining WORD with 50\% of HCC-TACE-SEG. 
One possible explanation is that WORD contains 80 samples, whereas HCC-TACE-SEG contains 105 samples, and the 50\% HCC-TACE-SEG setting may provide a better balance between diverse organ samples and liver pathology samples. 
This suggests that both diversity and semantic balance are important for effective server-side distillation.
Overall, CA-MMDS can operate with different proxy datasets, but proxy characteristics, including anatomical coverage, pathology patterns, data scale, and semantic balance, influence distillation quality. 
Although reliability-weighted aggregation reduces the influence of less compatible local models during pseudo-label aggregation, selecting suitable proxy data remains an important open problem. Future work will explore proxy data selection and distillation strategies that reduce the dependence on highly matched proxy datasets.

\input{exp/personal_FL}
\input{exp/different_backbone}
\input{exp/out_of_distribution}

\input{exp/hyparam_analysis}

\subsection{Analysis of Personalized Inference}
CA-MMDS enables personalized inference by allowing each client to use both its local model and the distilled global model. 
Figure~\ref{fig5: personal_FL} compares the class-wise performance of each client before and after five-stage FCL. 
The local model preserves client-specific expertise, while the global model provides knowledge aggregated from the evolving federation. 
For personalized inference, local predictions are prioritized for classes in the client's label set, while global predictions support segmentation of classes unavailable locally.

\subsection{Analysis of Heterogeneous Client Architectures}
As CA-MMDS distills prediction-level outputs rather than aggregating model parameters, it is compatible with heterogeneous client architectures.
To evaluate this property, we repeat the five-stage FCL process using different segmentation backbones across clients.
Specifically, Swin UNETR~\cite{hatamizadeh2021swin} is used for the second and fourth clients, while 3D U-Net~\cite{cciccek20163d} is used for the remaining clients.
Figure~\ref{fig6: achriteture} compares the final global model performance under homogeneous and heterogeneous client architectures.
The results show that CA-MMDS can aggregate knowledge from client models with different architectures and achieve performance comparable to the homogeneous setting.

\subsection{Out-of-Distribution Evaluation}
We further evaluate out-of-distribution generalization on the BTCV dataset~\cite{landman2015miccai}. 
Figure~\ref{fig7: ood} reports the Dice and ASSD performance of the final global model on BTCV, with centralized training included as the upper bound. The ``Portal Vein and Splenic Vein'' class is excluded because it is not part of the federation label space. 
The results show that CA-MMDS achieves performance comparable to centralized training on most classes, indicating that the distilled global model generalizes well to unseen datasets.

\subsection{Hyperparameter Analysis}
We analyze the sensitivity of CA-MMDS to the three hyperparameters used in reliability-weighted aggregation: the freshness decay factor $\gamma$, the confidence temperature $\beta$, and the domain compatibility temperature $\eta$. 
We vary one hyperparameter at a time while keeping the others fixed to their default values, and evaluate the final global model after five-stage FCL. 
As shown in Figure~\ref{fig:hyperparameter}, CA-MMDS remains stable within a reasonable range of each hyperparameter. 
The default setting achieves the best or near-best performance, indicating that the proposed reliability-weighted aggregation is robust to moderate hyperparameter variations.

%% file: section/5_1_experiment_setup.tex
\subsection{Datasets}
We conduct experiments on publicly available 3D abdominal CT segmentation datasets.
Five datasets are used to construct the FCL scenario: LiTS~\cite{bilic2023liver}, KiTS~\cite{heller2019kits19}, MSD Pancreas~\cite{antonelli2022medical}, MSD Spleen~\cite{antonelli2022medical}, and AMOS~\cite{ji2022amos}.
Together, these datasets provide annotations for 19 abdominal organs and associated pathologies, with partially overlapping and dataset-specific label spaces.
Following~\cite{xu2023federated,kim2024federated}, BTCV~\cite{landman2015miccai} is used as an external out-of-distribution test set and is excluded from federation training.
WORD~\cite{luo2021word} serves as the unlabeled proxy dataset for server-side distillation.

\subsection{Evaluation Metrics}
We evaluate segmentation performance using the Dice similarity coefficient (Dice) and average symmetric surface distance (ASSD), which measure region overlap and boundary accuracy, respectively.
We report the communication and computation costs at each federation update to assess efficiency.

\subsection{Task Setup}
To simulate multi-site medical collaborations, we treat each dataset as a client and use all available annotations. 
This naturally creates heterogeneous label spaces, where clients may share anatomical targets while also containing client-specific organs or pathologies.
We construct a five-stage FCL segmentation scenario with both new and evolving clients. 
The federation starts with MSD Pancreas and MSD Spleen at stage 1, followed by the sequential introduction of KiTS, LiTS, and AMOS in stages 2--4. 
To simulate client evolution, AMOS is split into two subsets with different annotation targets, allowing its label space to expand at stage 5. 
Thus, stages 2--4 represent newly added clients, while stage 5 represents an updated client.
At each FCL stage, evaluation is performed on the accumulated test sets of all clients seen so far.

\subsection{Implementation Details}
CA-MMDS is implemented in PyTorch using CLIP-DrivenSeg~\cite{liu2023clip} as the segmentation backbone. 
All models are trained with AdamW and a warm-up learning rate scheduler. 
The initial learning rate is $4 \times 10^{-4}$, with $\beta_1=0.9$ and weight decay $1 \times 10^{-5}$. 
MAPCR-based baselines are trained for 1000 communication rounds with one local epoch per round. 
For server-side distillation methods, the global model is trained for 1000 epochs on the distillation dataset, and the checkpoint with the best validation Dice is selected.

%% file: exp/main_compare_tab.tex
\definecolor{stepblue}{RGB}{230,242,250} 

\begin{table*}[t]
\centering
\caption{
    Main comparison after 5-stage FCL segmentation with evolving clients and label spaces.
    Dice (\%) and ASSD (mm) measure segmentation performance, while cumulative communication and computation costs measure training efficiency.
    Here, $1\times$ denotes the computation required to train one client model to convergence, assumed comparable across clients.
}
\footnotesize
\setlength{\tabcolsep}{4pt}
\renewcommand{\arraystretch}{1.0}
\begin{tabular}{l|c|cc|cc|cc|cc}
\hline
\multirow{2}{*}{Method}
& \multirow{2}{*}{\makecell{Supports offline-\\client training?}}
& \multicolumn{2}{c|}{Training Efficiency}
& \multicolumn{2}{c|}{Overall}
& \multicolumn{2}{c|}{Organ}
& \multicolumn{2}{c}{Pathology} \\
&
& Comm. $\downarrow$ & Comp. $\downarrow$
& Dice $\uparrow$ & ASSD $\downarrow$
& Dice $\uparrow$ & ASSD $\downarrow$
& Dice $\uparrow$ & ASSD $\downarrow$ \\
\hline
Centralized Training (upper bound) & -- & -- & -- & 78.99 & 4.37 & 84.73 & 3.27 & 57.46 & 8.49 \\
FedAvg~\cite{mcmahan2017communication} & \cross & $38000$ & $19\times$ & 77.66 & 4.38 & 84.24 & 2.71 & 53.00 & 10.61 \\
\hline
FedAvg~\cite{mcmahan2017communication} (1 round) & \cross & $38$ & $0.19\times$ & 0.00 & 271.06 & 0.00 & 271.06 & 0.00 & -- \\
FedProx~\cite{li2020federated} (1 round) & \cross & $38$ & $0.19\times$ & 0.00 & 257.27 & 0.00 & 257.27 & 0.00 & -- \\
FedAvg~\cite{mcmahan2017communication} (10 round) & \cross & $380$ & $1.9\times$ & 5.08 & 93.63 & 6.16 & 96.61 & 1.06 & 82.46 \\
FedProx~\cite{li2020federated} (10 round) & \cross & $380$ & $1.9\times$ & 4.17 & 102.74 & 4.12 & 113.33 & 4.38 & 63.06 \\
Kim et al.-style KD~\cite{kim2025communication} (adapted) & \tick & $25$ & $11\times + 3\times_{\mathrm{Synth.}}$ & 65.85 & 17.75 & 78.41 & 9.15 & 18.70 & 50.14 \\
Uniform MMDS & \tick & $25$ & $11\times$ & 66.27 & 12.16 & 78.45 & 7.63 & 20.60 & 29.13 \\
\rowcolor{stepblue} CA-MMDS (ours) & \tick & $25$ & $11\times$ & \textbf{67.49} & \textbf{9.83} & \textbf{79.73} & \textbf{5.05} & \textbf{21.55} & \textbf{27.76} \\
\hline
\end{tabular}
\label{tab:main_comparison}
\vspace{-5pt}
\end{table*}

%% file: exp/main_compare.tex
\begin{figure*}[t]
    \centering
    \includegraphics[width=1.0\textwidth]{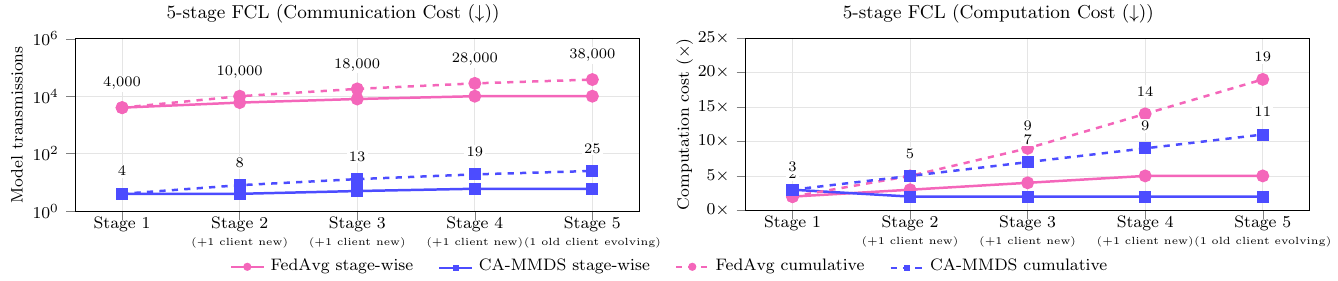}
    \caption{
    Communication and computation costs of MAPCR-type FedAvg and CA-MMDS over five FCL stages.
    }
	\label{fig3: communication}
    \vspace{-5pt}
\end{figure*}

\begin{figure*}[h]
    \centering
    \includegraphics[width=\textwidth]{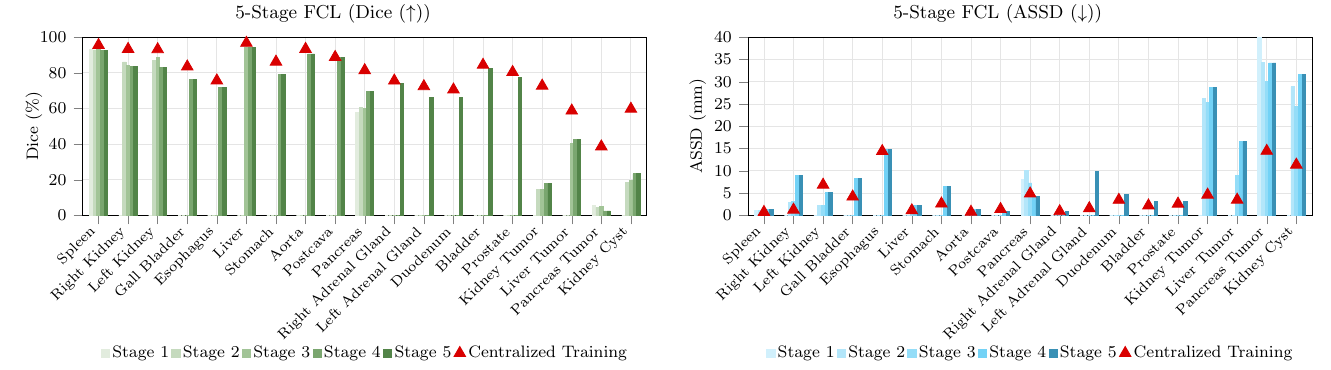}
    \caption{
        Class-wise Dice and ASSD performance of the proposed CA-MMDS global model on 5-stage FCL segmentation. 
    }
	\label{fig4: accuracy}
    \vspace{-10pt}
\end{figure*}

%% file: exp/ablation_exp.tex
\begin{table*}[t]
\centering
\caption{
    Ablation study of CA-MMDS.
    Segmentation performance is reported using Dice (\%) and ASSD (mm).
}
\footnotesize
\setlength{\tabcolsep}{5pt}
\renewcommand{\arraystretch}{1.0}
\begin{tabular}{l|cc|cc|cc}
\hline
\multirow{2}{*}{Method} & \multicolumn{2}{c|}{Overall} & \multicolumn{2}{c|}{Organ} & \multicolumn{2}{c}{Pathology} \\
& Dice $\uparrow$ & ASSD $\downarrow$ & Dice $\uparrow$ & ASSD $\downarrow$ & Dice $\uparrow$ & ASSD $\downarrow$ \\
\hline
Uniform MMDS & 66.27 & 12.16 & 78.45 & 7.63 & 20.60 & 29.13 \\
CA-MMDS w/o model freshness ($r_{m,t}$) & 66.90 & \textbf{9.65} & 79.13 & 6.15 & 21.03 & \textbf{22.77} \\
CA-MMDS w/o prediction confidence ($q_{m,c}(\mathbf{X})$) & \underline{67.15} & 10.55 & 79.15 & 5.87 & \textbf{22.15} & 28.12 \\ 
CA-MMDS w/o domain compatibility ($b_m(\mathbf{X})$) & 67.14 & 11.38 & \underline{79.41} & \underline{5.30} & 21.10 & 34.17 \\
CA-MMDS & \textbf{67.49} & \underline{9.83} & \textbf{79.73} & \textbf{5.05} & \underline{21.55} & \underline{27.76} \\
\hline
\end{tabular}
\label{tab:ablation_reliability}
\vspace{-5pt}
\end{table*}

%% file: exp/proxy_dataset_analysis_tab.tex
\begin{table}[t]
\centering
\caption{ 
Analysis of CA-MMDS with different distillation datasets: WORD~\cite{luo2021word} and HCC-TACE-SEG (HCC)~\cite{moawad2023multimodality}.
}
\label{tab:proxy_sensitivity}
\setlength{\tabcolsep}{1.8pt}
\renewcommand{\arraystretch}{1.0}
\begin{tabular}{l|cc|cc|cc}
\hline
\multirow{2}{*}{Proxy dataset} & \multicolumn{2}{c|}{Overall} & \multicolumn{2}{c|}{Organ} & \multicolumn{2}{c}{Pathology} \\
 & Dice $\uparrow$ & ASSD $\downarrow$ & Dice $\uparrow$ & ASSD $\downarrow$ & Dice $\uparrow$ & ASSD $\downarrow$ \\
\hline
WORD 100\% & 67.49 & 9.83 & 79.73 & 5.05  & 21.55 & 27.76 \\
WORD 50\% & 62.90 & 12.81 & 76.44 & 6.02 & 11.97 & 38.26 \\ 
WORD 25\% & 58.56 & 15.17 & 70.95 & 8.42 & 12.10 & 40.51 \\
\hline
HCC 100\% & 62.45 & 14.23 & 72.56 & 11.42 & 24.54 & 24.77 \\
HCC 50\% & 56.74 & 23.17 & 67.08 & 16.99 & 17.97 & 46.34 \\
\hline
WORD + HCC 100\% & \underline{70.09} & \underline{8.76} & \underline{80.93} & \underline{5.15} & \textbf{29.47} & \textbf{22.26} \\
WORD + HCC 50\% & \textbf{70.35} & \textbf{7.76} & \textbf{81.61} & \textbf{3.35} & \underline{28.12} & \underline{24.28} \\
\hline
\end{tabular}
\vspace{-5pt}
\end{table}

%% file: exp/personal_FL.tex
\begin{figure*}[t]
    \centering
    \includegraphics[width=\textwidth]{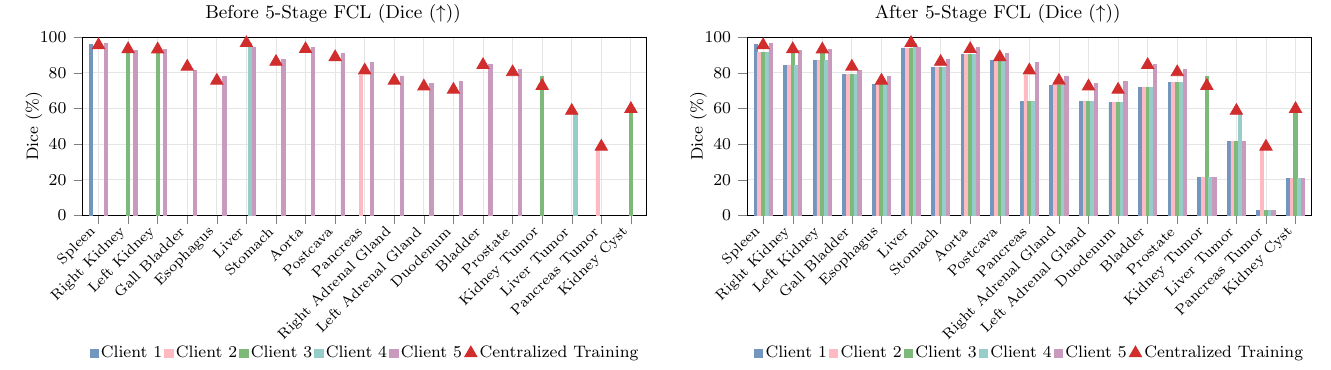}
    \caption{
    Dice performance of each client before and after the 5-stage FCL.
    The local model complements the global model with client-specific knowledge for personalized inference.
    }
    \vspace{-10pt}
	\label{fig5: personal_FL}
\end{figure*}

%% file: exp/different_backbone.tex
\begin{figure}[t]
  \centering
  \includegraphics[width=\linewidth]{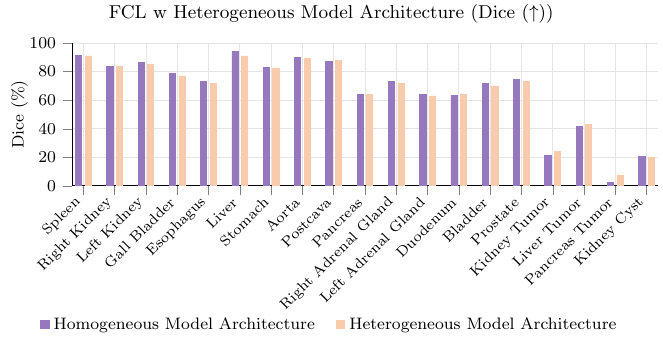} 
    \caption{
        Dice performance of the global model under homogeneous and heterogeneous model architectures.
    }
    \vspace{-10pt}
	\label{fig6: achriteture}
\end{figure}

%% file: exp/out_of_distribution.tex
\begin{figure}[t]
    \centering
    \includegraphics[width=\linewidth]{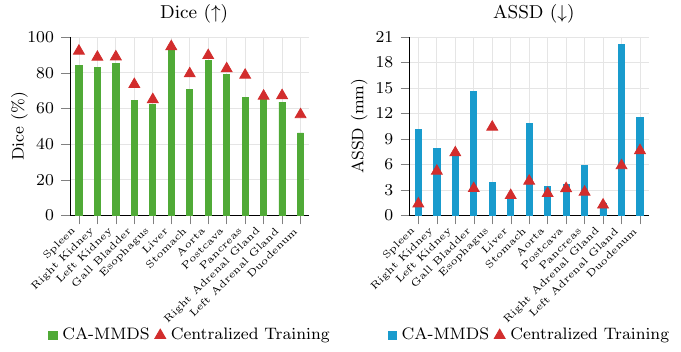} 
    \caption{
    Out-of-distribution performance of the global model after 5-stage FCL, evaluated on the BTCV dataset \cite{landman2015miccai}.
    }
    \vspace{-10pt}
	\label{fig7: ood}
\end{figure}

%% file: exp/hyparam_analysis.tex
\begin{figure}[t]
\centering
\includegraphics[width=\linewidth]{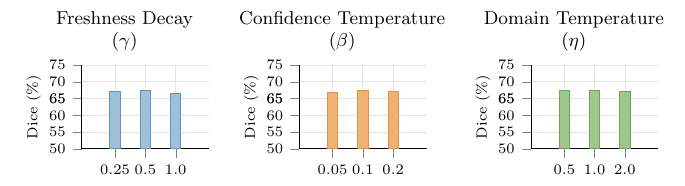}
\caption{
Hyperparameter sensitivity analysis of CA-MMDS.
}
\label{fig:hyperparameter}
\vspace{-10pt}
\end{figure}

%% file: section/6_conclusion.tex
\section{Conclusion}
This paper studies asynchronous federated continual learning for multi-class 3D segmentation, where clients and label spaces may evolve over time. 
We propose CA-MMDS, a continual archive-based multiple-model distillation framework that maintains a server-side archive of client models and updates the global model through reliability-weighted distillation. 
By considering class ownership, model freshness, prediction confidence, and domain compatibility, CA-MMDS effectively integrates knowledge from evolving clients without repeatedly synchronizing all participants. 
Experiments on 3D abdominal CT segmentation show that CA-MMDS substantially reduces communication and computation costs while achieving competitive segmentation performance and supporting asynchronous client updates.

%% file: section/7_supplymentary.tex
\section{Supplementary Material}

\subsection{Notation Summary}
\label{app:notation}

Table~\ref{tab:notation_summary} summarizes the main symbols used in the method section.

\subsection{Dataset}
\label{sup_dataset}

In this section, we introduce the 3D abdominal CT segmentation datasets used in our experiments. 
Since different datasets provide annotations for varying classes, with some overlap between them, we followed the approach outlined in \cite{liu2023clip} to unify the label indices across all datasets.
Overall, we focus on segmenting 19 abdominal organs and pathological structures. 
As many datasets do not provide ground truth annotations for their test data, following \cite{liu2023clip}, we reorganize the available training sets, splitting them into training, validation, and testing subsets.
The detailed information for each dataset is provided below, and a summary is presented in Table \ref{sub_tab1: dataset}, including the number of training samples, the number of class labels, and the label lists.

\noindent \textbf{BTCV}\cite{landman2015miccai} 
comprises 30 scans with volume size range from $[85 \sim 198] \times [512] \times [512]$ pixels in dimensions $D \times H \times W$, , where $D$ represents the depth, $H$ is the height, and $W$ is the width.
As the BTCV dataset is used for out-of-distribution performance evaluation, all its data are used for testing.
The in-plane pixel spacing in this dataset varies between 0.59mm to 0.98mm, and the slice thickness ranges from 2.50mm to 5.00mm.
The dataset provides pixel-wise annotations for 14 classes: spleen, right Kidney, left kidney, gall bladder, esophagus, liver, stomach, aorta, postcava, portal vein and splenic vein, pancreas, right adrenal gland, left adrenal gland, and duodenum.

\noindent \textbf{LiTS (2017)} \cite{bilic2023liver} 
comprises 131 scans with volume size range from $[74 \sim 987] \times [512] \times [512]$ pixels in dimensions $D \times H \times W$.
Out of the 131 scans, 95 are used for training, 16 for validation, and 20 for testing.
The in-plane pixel spacing in this dataset varies between 0.56mm and 1.00mm, and the slice thickness ranges from 0.70mm to 5.00mm.
The dataset provides annotations for 2 classes: liver and liver tumor.

\noindent \textbf{KiTS} \cite{heller2019kits19} 
comprises 210 scans with volume sizes ranging from $[29 \sim 1059] \times [512] \times [512,796]$ pixels in dimensions $D \times H \times W$.
Out of the 210 scans, 142 are used for training, 22 for validation, and 46 for testing.
The in-plane pixel spacing in this dataset varies between 0.44mm and 1.04mm, and the slice thickness ranges from 0.50mm to 5.00mm.
The dataset provides annotations for 4 classes: left kidney, right kidney, kidney tumor, and kidney crystal.

\noindent \textbf{MSD - Spleen} \cite{antonelli2022medical}
comprises 41 scans with volume size range from $[31 \sim 168] \times [512] \times [512]$ pixels in dimensions $D \times H \times W$.
Out of the 41 scans, 27 are used for training, 4 for validation, and 10 for testing.
The in-plane pixel spacing in this dataset varies between 0.61mm and 0.98mm, and the slice thickness ranges from 1.50mm to 8.00mm.
The dataset provides annotations for 1 class: spleen.

\noindent \textbf{MSD - Pancreas} \cite{antonelli2022medical}
comprises 281 scans with volume size range from $[37 \sim 751] \times [512] \times [512]$ pixels in dimensions $D \times H \times W$.
Out of the 281 scans, 198 are used for training, 29 for validation, and 54 for testing.
The in-plane pixel spacing in this dataset varies between 0.54mm and 0.98mm, and the slice thickness ranges from 0.63mm to 7.50mm.
The dataset provides annotations for 2 classes: pancreas and pancreas tumor.

\noindent \textbf{AMOS} \cite{ji2022amos} 
comprises 200 scans with volume size range from $[64 \sim 512] \times [60~768] \times [192 \sim 768]$ pixels in dimensions $D \times H \times W$.
Out of the 200 scans, 145 are used for training, 23 for validation, and 32 for testing.
The in-plane pixel spacing in this dataset varies between 0.45mm and 3.00mm, and the slice thickness ranges from 0.82mm to 5.00mm.
The dataset provides annotations for 15 classes: spleen, right kidney, left kidney, gall bladder, esophagus, liver, stomach, aorta, postcava, pancreas, right adrenal gland, left adrenal gland, duodenum, bladder, and prostate.

\input{table/dataset_table}

\input{table/notation_table}

\subsection{Visualization}
We present visualization results comparing the global model trained with our proposed method against local training and ground truth, shown in both 2D and 3D views. 
Figure \ref{fig8: 2D_visualization} and Figure \ref{fig9: 3D_visualization} display the visualization results in 2D (axial) and 3D views, respectively.
Each column represents a case from a different client dataset.
The middle five rows illustrate the segmentation results produced by the global model on all its client datasets after each stage of FCL learning. 
It can be observed that when new clients are joining or existing clients are evolving, the proposed method is flexible to adapt to the dynamic FCL scenarios and keep updating the global model to be able to segment an increasing number of objects of interest.

\input{exp/visualization}

%% file: table/dataset_table.tex
\begin{table}[t]
    \centering
    \scriptsize
    \caption{
    This table presents the dataset information used in the paper.
    To simulate real-world conditions, each dataset is treated as an individual client, with all available class annotations utilized. 
    Consequently, class distributions may vary between clients, and some classes may overlap across different clients, mirroring real-world scenarios.
    }
    \begin{tabular}{l|c|c|l}
    \hline
    \textbf{Datasets} & \textbf{\# Labels} & \textbf{\# Train Imgs} & \textbf{Annotated Structures} \\ \hline
    LiTS & 2 & 131 & Liver, Liver Tumor \\ \hline
    KiTS & 4 & 210 & \makecell[l]{Left Kidney (LKid), \\ 
                                    Right Kidney (RKid), \\ 
                                    Kidney Tumor, Kidney Crystal} \\ \hline
    MSD-Spleen & 1 & 41 & Spleen \\ \hline
    MSD-Pancreas & 2 & 281 & Pancreas, Pancreas Tumor \\ \hline
    AMOS22 & 15 & 200 & \makecell[l]{Spleen, RKid, LKid, \\
                                        Gall bladder (Gall), \\
                                        Esophagus, Liver, Stomach, \\
                                        Aorta, IVC, Pancreas, \\
                                        Right adrenal gland (RAG), \\
                                        Left adrenal gland (LAG), \\
                                        Duodenum, Bladder, Prostate} \\ \hline
    BTCV & 14 & 30 & \makecell[l]{Spleen, RKid, LKid, Gall, \\
                                    Esophagus, Liver, Stomach, \\
                                    Aorta, IVC, R\&S Veins, \\
                                    Pancreas, RAG, \\
                                    LAG, Duodenum} \\ \hline
    \end{tabular}
\label{sub_tab1: dataset}
\end{table}

%% file: table/notation_table.tex
\begin{table*}[t]
\centering
\small
\renewcommand{\arraystretch}{1.1}
\caption{Summary of the main notation used in the proposed CA-MMDS framework.}
\label{tab:notation_summary}
\begin{tabular}{p{0.22\linewidth}p{0.72\linewidth}}
\hline
\textbf{Symbol} & \textbf{Description} \\
\hline

\multicolumn{2}{l}{\textbf{Federated continual learning setting}} \\
$t$ & Current federation stage. \\
$\tau$ & Historical upload stage of an archived local model. \\
$k$ & Client index. \\
$m$ & Index of an archived local model in $\mathcal{A}_t$. \\
$c$ & Class index. \\
$\mathcal{S}_t$ & Set of accumulated clients at stage $t$. \\
$\mathcal{U}_t$ & Set of newly added or updated clients that perform local training at stage $t$. \\
$\mathcal{C}_t$ & Cumulative label set at stage $t$. \\
$\mathcal{C}_{k,t}$ & Label set annotated by client $k$ at stage $t$. \\
$\mathcal{D}_{k,t}$ & Private annotated dataset of client $k$ at stage $t$. \\
$\mathcal{D}_p$ & Public unlabeled proxy dataset used for server-side distillation. \\

\multicolumn{2}{l}{\textbf{Input, output, and local training}} \\
$\mathbf{X}$ & Input 3D volume. \\
$C_{\mathrm{in}}$ & Number of input imaging channels. \\
$H,W,Z$ & Spatial height, width, and depth of a 3D volume. \\
$\Omega_{\mathbf{X}}$ & Spatial voxel domain of $\mathbf{X}$. \\
$N$ & Number of voxels in $\Omega_{\mathbf{X}}$, where $N=HWZ$. \\
$n$ & Voxel index in $\Omega_{\mathbf{X}}$. \\
$f^L_{k,t}$ & Local model trained by client $k$ at stage $t$. \\
$f^G_t$ & Global model maintained by the server at stage $t$. \\
$\mathbf{P}^L_{k,t}$ & Probability map predicted by local model $f^L_{k,t}$. \\
$\mathbf{P}^G_t$ & Probability map predicted by global model $f^G_t$. \\
$\mathbf{Y}_{k,t,c}$ & Binary ground-truth segmentation mask for class $c$ on client $k$ at stage $t$. \\
$\mathcal{L}^{L}_{k,t}$ & Local training objective for client $k$ at stage $t$. \\
$\mathcal{L}^{G}_{\mathrm{dist}}$ & Global distillation objective on the proxy dataset. \\

\multicolumn{2}{l}{\textbf{Model archive}} \\
$\mathcal{A}_t$ & Server-side archive of local models available at stage $t$. \\
$\mathcal{H}_{k,t}$ & Set of historical upload stages of client $k$ up to stage $t$. \\

\multicolumn{2}{l}{\textbf{Reliability-weighted aggregation}} \\
$o_{k,\tau,c}$ & Class ownership indicator showing whether $f^L_{k,\tau}$ was trained for class $c$. \\
$r_{k,\tau,t}$ & Freshness score of model $f^L_{k,\tau}$ at stage $t$. \\
$\gamma$ & Decay coefficient controlling freshness down-weighting. \\
$E_{k,\tau,c}(\mathbf{X})$ & Voxel-wise binary entropy of the prediction for class $c$ on proxy volume $\mathbf{X}$. \\
$q_{k,\tau,c}(\mathbf{X})$ & Prediction confidence score derived from entropy. \\
$\beta$ & Temperature parameter for converting entropy into confidence. \\
$\mathcal{B}_{k,\tau}$ & BN statistics stored in archived local model $f^L_{k,\tau}$. \\
$\mathcal{J}_{\mathrm{BN}}$ & Set of BN layers. \\
$j$ & BN layer index. \\
$\boldsymbol{\mu}^{j}_{k,\tau}, \boldsymbol{\sigma}^{j}_{k,\tau}$ & Running mean and standard deviation stored in the $j$-th BN layer of $f^L_{k,\tau}$. \\
$\widehat{\boldsymbol{\mu}}^{j}_{k,\tau}(\mathbf{X}), \widehat{\boldsymbol{\sigma}}^{j}_{k,\tau}(\mathbf{X})$ & Activation mean and standard deviation computed from proxy volume $\mathbf{X}$. \\
$d_{k,\tau}(\mathbf{X})$ & BN compatibility distance between proxy activations and stored BN statistics. \\
$b_{k,\tau}(\mathbf{X})$ & Domain compatibility score. \\
$\eta$ & Temperature parameter for converting BN distance into domain compatibility. \\
$\alpha_{m,c}(\mathbf{X})$ & Reliability weight of archived model $m$ for class $c$ on proxy volume $\mathbf{X}$. \\
$\widetilde{\mathbf{Y}}_{t,c}(\mathbf{X})$ & Aggregated pseudo-label for class $c$ at stage $t$. \\
$\epsilon$ & Small constant for numerical stability. \\

\multicolumn{2}{l}{\textbf{Efficiency analysis}} \\
$E$ & Number of communication rounds used by MAPCR. \\
$O$ & Computation required to train one local or global model to convergence. \\

\hline
\end{tabular}
\end{table*}

%% file: exp/visualization.tex
\begin{figure*}[b]
    \centering
    \includegraphics[width=0.85\linewidth, trim={0.0cm 0.0cm 0.0cm 0.0cm},clip]{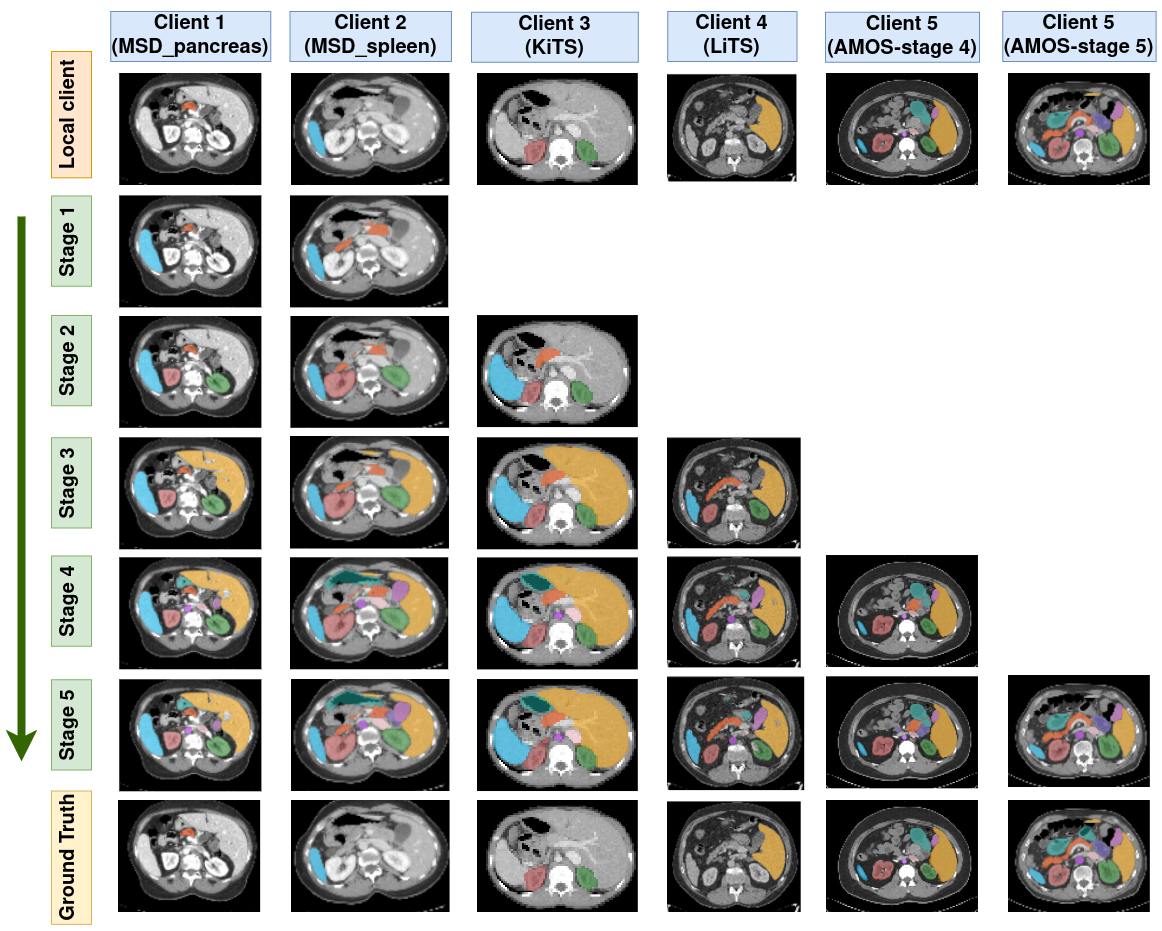} 
    \caption{2D visualizations of segmentation results produced by the proposed CA-MMDS method under the 5-stage FCL.
    Each column displays a case from either a new client joining the federation or an updated client. 
    As the federation expands and existing clients evolve, the global model progressively gains the capability to segment new objects introduced by these clients.
    }
	\label{fig8: 2D_visualization}
\end{figure*}

\begin{figure*}[b]
    \centering
    \includegraphics[width=0.85\linewidth, trim={0.0cm 0.0cm 0.0cm 0.0cm},clip]{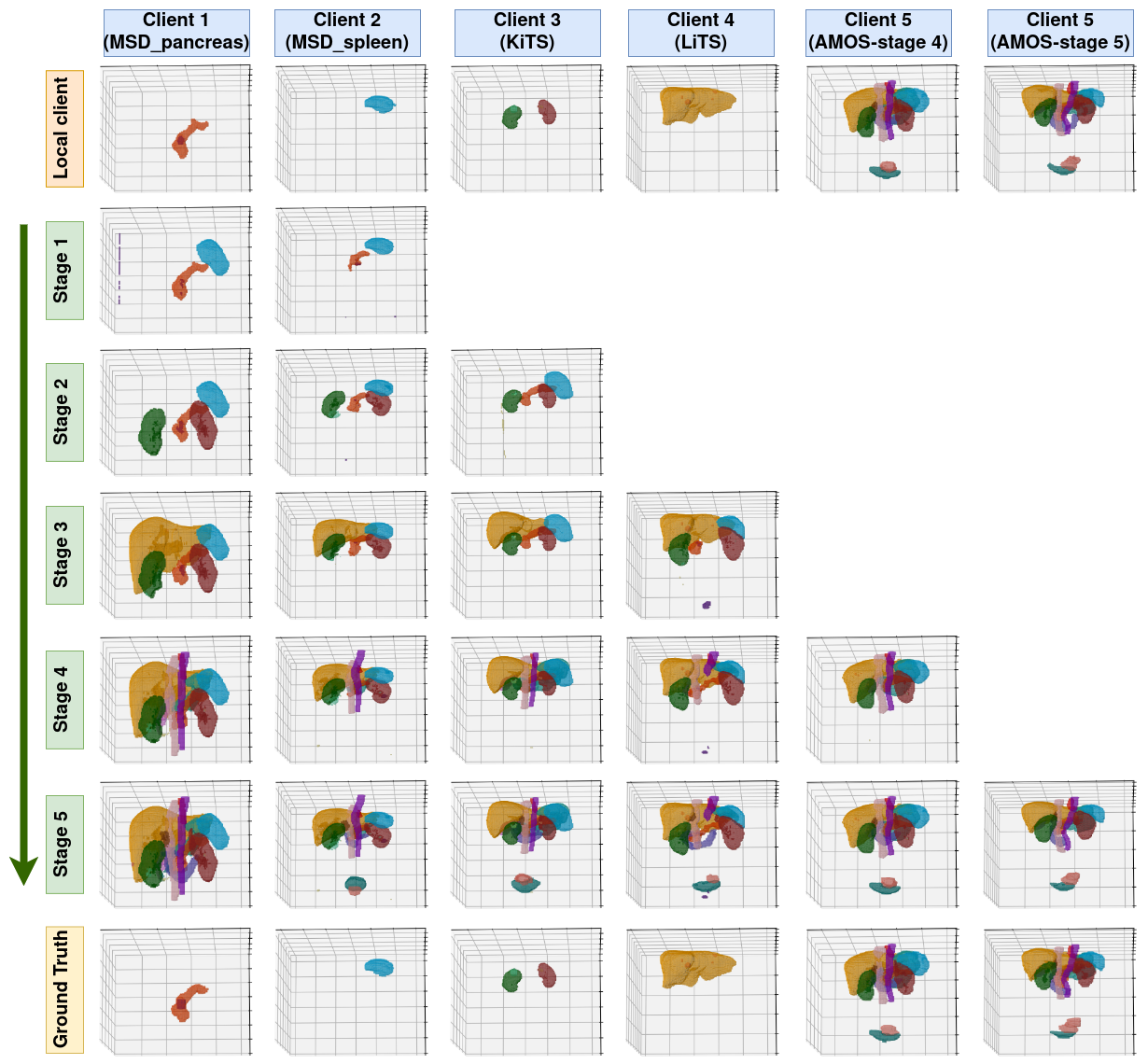} 
    \caption{3D visualizations of segmentation results produced by the proposed CA-MMDS method under the 5-stage FCL.
    }
	\label{fig9: 3D_visualization}
\end{figure*}